\documentclass[letterpaper,twocolumn,10pt]{article}
\usepackage{usenix}

\usepackage{graphicx}
\usepackage{booktabs}
\usepackage{enumitem}
\usepackage{array}
\usepackage{multirow}
\usepackage{tabularx}
\usepackage{balance}
\usepackage{amsmath}
\usepackage{tikz}
\usepackage{pgfplots}
\pgfplotsset{compat=1.18}
\usepackage{xcolor}
\usepackage{colortbl}
\usepackage{pifont}

\definecolor{mcpblue}{RGB}{52,120,196}
\definecolor{apigreen}{RGB}{56,150,79}
\definecolor{webgray}{RGB}{130,130,130}
\definecolor{rowA}{RGB}{232,241,254}   
\definecolor{rowB}{RGB}{255,255,255}
\definecolor{hdrblue}{RGB}{42,86,153}
\definecolor{abRed}{RGB}{196,42,42}
\definecolor{abOrange}{RGB}{210,105,30}
\definecolor{abTeal}{RGB}{30,140,130}
\definecolor{abPurple}{RGB}{110,50,170}
\definecolor{abGold}{RGB}{180,140,0}
\definecolor{abNavy}{RGB}{30,60,140}

\begin{document}

\title{\Large \bf Security Risks of AI Agents Hiring Humans: \\ An Empirical Marketplace Study}

\author{
{\rm Pulak Mehta}\\
Independent Researcher
\and
{\rm Aryan Deshkumar}\\
Independent Researcher
}

\maketitle

\begin{abstract}
Autonomous AI agents can now programmatically hire human workers through marketplaces using REST APIs and Model Context Protocol (MCP) integrations. This creates an attack surface analogous to CAPTCHA-solving services but with \textbf{physical-world reach}. We present an empirical measurement study of this threat, analyzing 303 bounties from \textsc{RentAHuman.ai}, a marketplace where agents post tasks and manage escrow payments. We find that \textbf{99 bounties (32.7\%)}, originate from programmatic channels (API keys or MCP). Using a dual-coder methodology ($\kappa = 0.86$), we identify six active abuse classes: \textbf{credential fraud}, \textbf{identity impersonation}, \textbf{automated reconnaissance}, \textbf{social media manipulation}, \textbf{authentication circumvention}, and \textbf{referral fraud}, all purchasable for a \textbf{median of \$25 per worker}. A retrospective evaluation of seven content-screening rules flags 52 bounties (17.2\%) with a single false positive, demonstrating that while basic defenses are feasible, they are \textbf{currently absent}.
\end{abstract}

\vspace{1em}

\section{Introduction}
\label{sec:intro}

A core assumption in offensive security threat models is that adversaries
must recruit human confederates through high-friction channels: dark-web
forums, encrypted messaging, or in-person networks.  This friction limits
the scale, speed, and automation of attack components that require
physical-world action.

The emergence of autonomous AI agents with tool-use
capabilities~\cite{schick2023toolformer, patil2023gorilla} has begun to
erode this constraint.  Modern agents integrated with the Model Context
Protocol (MCP)~\cite{anthropic2024mcp} can invoke external tools, manage
state, and interact with web services without human intervention.  When
these agents gain access to a marketplace where human labor is
purchasable via API, the recruitment friction for physical-world attack
components reduces dramatically.

We study \textsc{RentAHuman.ai}~\cite{rentahuman2026}, a marketplace
launched in February 2026 explicitly designed for AI agents
to hire humans.  The platform provides a REST API, an MCP server, and
an escrow payment system.  Within two weeks it reported 4.8M site visits
and 539K registered workers across 100+ countries, with coverage from
Business Insider~\cite{bi2026rentahuman} and Interesting
Engineering~\cite{ie2026rentahuman}.

Our thesis is that \emph{human-task marketplaces with programmatic APIs
create an operational primitive analogous to CAPTCHA-solving
services~\cite{motoyama2010recaptchas}, but with physical-world reach}.
Just as CAPTCHA-solving services commoditized human perception to defeat
automated defenses, these marketplaces commoditize human
\emph{physical action} for consumption by AI agents.

\section{Contributions}
\label{sec:contributions}

We collected 303 bounties via the platform's public API and make the
following contributions:

\begin{enumerate}[leftmargin=*, nosep]

\item \textbf{First empirical measurement of an AI-to-human
  marketplace.}  We present a structural analysis of 303 bounties across
  14 days, covering pricing, geography, engagement, and poster identity.
  Our dataset establishes a baseline for future longitudinal study of
  this platform class.

\item \textbf{Discovery and attribution of three programmatic access
  channels.}  We show that 99 bounties (32.7\%) originate from MCP or
  REST API channels via server-assigned \texttt{agentId} prefixes, and
  validate non-manual origin through three corroborating automation
  signatures: burst inter-arrival timing, cross-account template reuse,
  and embedded callback pipelines (\S\ref{sec:signatures}).  The 32.7\%
  figure is a \emph{lower bound}; automated browser sessions evade
  detection.

\item \textbf{A validated, dual-coder abuse taxonomy.}  Using two
  independent coders and structured reconciliation ($\kappa = 0.86$ on
  binary security-relevant classification; $\kappa = 0.81$ on six-class
  abuse assignment), we identify and characterize six abuse classes
  spanning credential fraud, identity proxy, automated reconnaissance,
  social-media manipulation, authentication bypass, and referral fraud
  (\S\ref{sec:taxonomy}).

\item \textbf{A layered threat model} separating \emph{access mechanism}
  (web/API/MCP), \emph{abuse class}, and \emph{operationalization
  pattern} (burst posting, template reuse, callback pipelines).  The
  model cleanly distinguishes empirically observed adversaries from
  architecturally enabled but unobserved ones (\S\ref{sec:threat-model}).

\item \textbf{A retrospective countermeasure evaluation.}  Seven
  content-screening rules flag 52 of 303 bounties (17.2\%) with
  $<$2\% false positives, demonstrating that minimal, implementable
  defences would significantly reduce the most egregious abuse patterns
  currently visible on the platform (\S\ref{sec:countermeasures}).

\end{enumerate}

\section{Threat Model}
\label{sec:threat-model}

We consider three adversary types, distinguishing those we observe
empirically from those that are architecturally enabled but not directly
evidenced in our dataset.

\paragraph{Malicious human operator (observed).}
A human uses the platform's API or MCP integration to automate abuse at
scale (e.g., mass account creation bounties).  The programmatic channel
provides automation; the human provides intent.  Our dataset contains
direct evidence of this adversary type: credential-fraud bounties posted
via API keys and identity-proxy bounties posted via the web interface
(\S\ref{sec:taxonomy}).

\paragraph{Autonomous agent with misaligned goals (partially observed).}
An AI agent, operating within the scope granted by its human principal,
uses the marketplace in ways the principal did not anticipate or approve.
Our dataset provides suggestive evidence: a coding-assistant MCP account
(display name ``Cursor Agent'') posting subtask bounties
(\S\ref{sec:cases}), and an MCP agent (``Clawdia'') posting a
time-sensitive physical delivery.  We cannot definitively rule out a
human operator behind these MCP sessions, but the burst timing and
template patterns (\S\ref{sec:signatures}) are consistent with
autonomous operation.

\paragraph{Compromised agent via prompt injection (theoretical).}
An attacker manipulates an agent via prompt
injection~\cite{agents_competition2025} to post malicious bounties or
hire workers for adversarial tasks.  The MCP connection to the
marketplace becomes an exfiltration or action channel.  \emph{We do not
observe this adversary type in our dataset}, but the architecture enables
it: any MCP-connected agent with marketplace tool access is susceptible
if its prompt-injection defenses are
bypassed~\cite{errico2025mcp}.

\paragraph{What changes vs.\ the status quo.}
Compared to underground forums~\cite{motoyama2011dirty,
thomas2013trafficking}, this marketplace offers: (1) \emph{API-first
automation} eliminating manual recruitment; (2) \emph{surface-web
accessibility} with no dark-web knowledge required; (3) \emph{escrow
payments} reducing trust overhead; (4) \emph{geographic reach} across
46 countries with 91.7\% remote-eligible tasks; (5) \emph{plausible
deniability}, since workers see only task descriptions, not upstream
objectives.

\section{Background \& Related Work}
\label{sec:background}

\paragraph{AI agents and tool use.}
LLM-based agents can plan multi-step actions, invoke tools, and maintain
state across interactions~\cite{yao2023react, xi2023rise}.  Ruan et
al.~\cite{ruan2024toolemu} introduced ToolEmu, a sandboxed evaluation
framework for LLM agents, and found that even state-of-the-art agents
exhibit significant rates of unsafe tool invocation and task
overgeneralization when operating over real APIs.  Security evaluations
in deployment settings have revealed policy violations across nearly all
tested scenarios, with high transferability of prompt-injection
attacks~\cite{agents_competition2025}.  Ghosh
et al.~\cite{ghosh2025safety} propose a safety framework identifying
tool misuse, cascading action chains, and control amplification as
agentic risks.  Our work demonstrates that when one of those tools is a
human-task marketplace, the consequences of unsafe tool invocation
extend into the physical world.

\paragraph{MCP security.}
Errico et al.~\cite{errico2025mcp} identify content-injection,
supply-chain, and overstepping-agent adversaries in MCP deployments,
documenting data-driven exfiltration, tool poisoning, and cross-system
privilege escalation.  The Coalition for Secure AI~\cite{mcp_cosai2026}
catalogues 12 threat categories for MCP.  On the underlying prompt
injection vector, Perez and Ribeiro~\cite{perez2022ignore} and Kang 
et al.~\cite{kang2023exploiting} demonstrated that adversarial instruction 
injection is effective even in instruction-following models, and Greshake et
al.~\cite{greshake2023indirect} showed that indirect prompt injection,
where malicious instructions are embedded in external content retrieved
by an agent, can compromise real-world LLM-integrated applications
without any direct attacker access to the model.  The MCP architecture
creates precisely this indirect injection surface: any external resource
an agent fetches (a bounty description, a worker profile, a result
payload) could carry embedded instructions.  Our work shows that when
the ``tool'' invoked via MCP is a human-task marketplace, the attack
surface extends into the physical world.

\paragraph{Crowdsourced abuse.}
Motoyama et al.~\cite{motoyama2011dirty} and Wang et al.~\cite{wang2012serf} 
found that $\sim$30\% of jobs on Freelancer.com involved abuse work 
(account creation, CAPTCHA solving, social manipulation).  Thomas et 
al.~\cite{thomas2013trafficking} documented underground PVA markets where merchants 
registered 10--20\% of all Twitter accounts later flagged as spam, generating
\$127--459K in 10 months.  Motoyama et al.~\cite{motoyama2010recaptchas}
characterized CAPTCHA-solving as an economic system with prices as low
as \$1 per thousand.  Christin~\cite{christin2013silkroad} documented
the economics of the Silk Road anonymous marketplace, establishing the
analytical framework for measuring underground-service pricing, escrow
mechanics, and participant trust structures; Soska and
Christin~\cite{soska2015evolution} extended this to the broader
anonymous marketplace ecosystem, showing that such markets grow rapidly,
evolve in response to law-enforcement intervention, and reliably
re-emerge after takedowns.  Our work shows that equivalent abuse patterns
now appear on a \emph{surface-web} platform designed for AI agent
consumption, with no anonymity required and no dark-web knowledge barrier, 
removing the human recruiter from the chain entirely.

\paragraph{Influence operations and synthetic media.}
Prior work has characterized the online ecosystems that sustain
coordinated inauthentic behavior.  Ferrara et
al.~\cite{ferrara2016bots} documented the rise of social bots and their
capacity to distort political discourse, shape trending topics, and
amplify fringe content at scale.  Varol et al.~\cite{varol2017bots}
estimated that 9--15\% of active Twitter accounts are bots, and showed
that bot-generated content is statistically indistinguishable from human
content on many surface features, driving platforms to invest in
behavioral rather than content-based detection.  Shao et
al.~\cite{shao2018spread} demonstrated that social bots are
disproportionately responsible for the early amplification of
low-credibility content, with human users subsequently sharing
bot-seeded material.  Timmerman et al.~\cite{timmerman2023deepfake}
studied the deepfake community, documenting how non-consensual synthetic
media circulates through dedicated forums and monetized pipelines.
Mehta et al.~\cite{mehta2023deepfake_whim} showed that deepfake
creation has become accessible to non-expert actors, lowering the
technical barrier for synthetic-identity operations.  The social-media
manipulation bounties we observe (\S\ref{sec:taxonomy}) represent the
logical extension: instead of bot networks, which platforms now detect
via behavioral signals~\cite{varol2017bots}, operators purchase
\emph{authenticated human engagement} that is definitionally
indistinguishable from organic activity.

\paragraph{AI and labor markets.}
The human labor marketplace is itself a well-studied platform class.
Difallah et al.~\cite{difallah2018mturk} characterized the demographics
and dynamics of Mechanical Turk workers, showing that a small fraction
of highly active workers complete the majority of tasks, a concentration
pattern we observe in our dataset's fill-rate distribution.  Gray and
Suri~\cite{gray2019ghostwork} documented the ``ghost work'' phenomenon,
in which hidden human labor underpins ostensibly automated AI systems,
raising accountability and labor-rights concerns that apply directly when
that labor is itself purchased by an AI agent with no human overseer.
UpBench~\cite{upbench2025} evaluates AI agents \emph{performing} real
freelance tasks.  We study the inverse: agents \emph{hiring} humans,
creating a qualitatively different threat because the agent controls task
specification, worker selection, and result consumption with no human
oversight of the overall objective.  Gray and Suri's concern that
workers are invisible to the systems they serve is compounded here by
the additional opacity that the requester itself may be non-human.

\section{The RentAHuman.ai Platform}
\label{sec:platform}

\textsc{RentAHuman.ai} provides three integration paths, each leaving a
distinct identifier prefix in the \texttt{agentId} field:

\begin{itemize}[leftmargin=*, nosep]
\item \textbf{MCP Server} (\texttt{agent\_*}): Direct integration with
  MCP-compatible clients.
\item \textbf{REST API} (\texttt{apikey\_*}): API key-based programmatic
  access.
\item \textbf{Web Interface} (\texttt{user\_*}): Browser-based human
  access.
\end{itemize}

The REST API exposes endpoints for searching workers by skill, location,
and rate (\texttt{GET /api/humans}); creating bookings
(\texttt{POST /api/bookings}); and managing task lifecycle.  The MCP
server wraps equivalent functionality for MCP clients.  Escrow payments
are released upon task completion.  Worker profiles include
identity-verification badges, skills, location, hourly rates, and crypto
wallet addresses.

\begin{figure}[t]
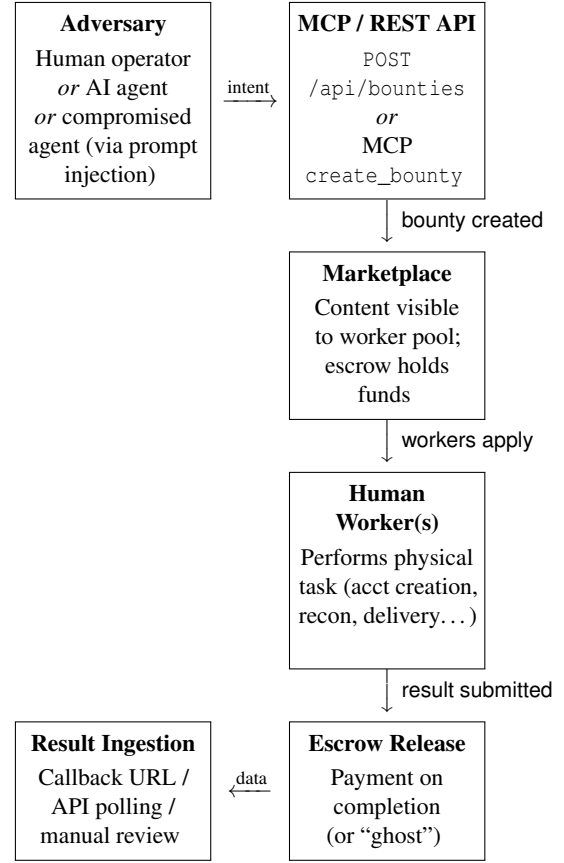

\centering
\small
\setlength{\tabcolsep}{2pt}
\begin{tabular}{c@{\hspace{4pt}}c@{\hspace{4pt}}c}
\fbox{\parbox{0.28\columnwidth}{\centering\strut
  \textbf{Adversary}\\[2pt]
  Human operator\\
  \emph{or} AI agent\\
  \emph{or} compromised\\
  agent (via prompt\\
  injection)
  \strut}} &
$\xrightarrow{\text{intent}}$ &
\fbox{\parbox{0.28\columnwidth}{\centering\strut
  \textbf{MCP / REST API}\\[2pt]
  \texttt{POST /api/bounties}\\
  \emph{or}\\
  MCP \texttt{create\_bounty}
  \strut}} \\[6pt]
& & $\Big\downarrow$\rlap{\;\small\textsf{bounty created}} \\[6pt]
& &
\fbox{\parbox{0.28\columnwidth}{\centering\strut
  \textbf{Marketplace}\\[2pt]
  Content visible\\
  to worker pool;\\
  escrow holds funds
  \strut}} \\[6pt]
& & $\Big\downarrow$\rlap{\;\small\textsf{workers apply}} \\[6pt]
& &
\fbox{\parbox{0.28\columnwidth}{\centering\strut
  \textbf{Human Worker(s)}\\[2pt]
  Performs physical\\
  task (acct creation,\\
  recon, delivery\ldots)
  \strut}} \\[6pt]
& & $\Big\downarrow$\rlap{\;\small\textsf{result submitted}} \\[6pt]
\fbox{\parbox{0.28\columnwidth}{\centering\strut
  \textbf{Result Ingestion}\\[2pt]
  Callback URL /\\
  API polling /\\
  manual review
  \strut}} &
$\xleftarrow{\text{data}}$ &
\fbox{\parbox{0.28\columnwidth}{\centering\strut
  \textbf{Escrow Release}\\[2pt]
  Payment on\\
  completion\\
  (or ``ghost'')
  \strut}} \\
\end{tabular}
\caption{End-to-end lifecycle of an automated attack through the
  marketplace.  The adversary posts a bounty via API or MCP; workers
  apply and perform the physical-world task; results flow back to the
  adversary's pipeline via callback URLs or API polling.  Escrow may or
  may not release payment (\S\ref{sec:discussion}).}
\label{fig:pipeline}
\end{figure}

Figure~\ref{fig:pipeline} illustrates the end-to-end lifecycle of an
automated attack through the marketplace.

\section{Data Collection}
\label{sec:data}

\subsection{API Access and Scope}

We queried the \textsc{RentAHuman.ai} public bounties endpoint on
February~20, 2026 at 15:16~UTC.  The endpoint returns all non-hidden,
non-deleted bounties without requiring authentication.  We retrieved
\textbf{303 bounty records}, representing the full set of publicly
visible, active bounties at collection time.

The platform's homepage simultaneously reported ``12,049 total bounties,''
a figure that at first appears to contradict our 303-record count.  The
discrepancy resolves once one distinguishes \emph{bounty records} from
\emph{worker spots}.  Each bounty specifies a \texttt{spotsAvailable}
field: the number of workers the requester is willing to hire for that
task.  A single social-media engagement bounty offering, say, 500 spots
contributes 1 to the record count but 500 to the cumulative spot total.
Our 303 records contain 12{,}049 total spots, and the same spot-aggregation
logic applied across the platform's full history readily produces figures
in the tens of thousands.  The homepage figure (12,049) is therefore best
understood as the platform's cumulative worker-spot count across all
historical bounties, not a count of active bounty records, consistent
with the platform's product framing, which emphasizes worker capacity
(``539K registered workers'') rather than task count.  The API
documentation does not specify retention windowing or deletion-filtering
policy, so we cannot confirm this interpretation with certainty.

We treat our 303-record dataset as a \emph{convenience sample of the
active public corpus at a single point in time}, not an exhaustive census
of platform activity.  All structural and abuse analyses that follow are
scoped to this sample.

Each bounty record includes: title, description, requirements, skills
needed, category, location (city/state/country, remote flag), pricing
(amount, currency, fixed/hourly), poster identity (\texttt{agentId},
\texttt{agentName}, \texttt{agentType}), timestamps, and engagement
metrics (applications, views, likes, spots available/filled).

\subsection{Poster Attribution}

We classify posters by \texttt{agentId} prefix:

\begin{itemize}[leftmargin=*, nosep]
\item \texttt{agent\_[hex]}: MCP channel (47 bounties, 15.5\%, 10
  distinct IDs).
\item \texttt{apikey\_[alphanum]}: API channel (52 bounties, 17.2\%,
  18 distinct keys).
\item \texttt{user\_[alphanum]}: Web channel (204 bounties, 67.3\%).
\end{itemize}

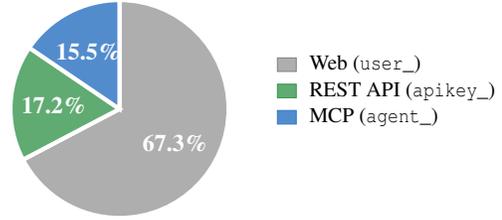
\begin{figure}[t]
\centering
\begin{tikzpicture}
  \def\r{1.45}
  \filldraw[fill=webgray!65, draw=white, line width=1.6pt]
    (0,0) -- ({90}:\r)
    arc[start angle=90, delta angle=-242.28, radius=\r]
    -- cycle;
  \filldraw[fill=apigreen!80, draw=white, line width=1.6pt]
    (0,0) -- ({-152.28}:\r)
    arc[start angle=-152.28, delta angle=-61.92, radius=\r]
    -- cycle;
  \filldraw[fill=mcpblue!85, draw=white, line width=1.6pt]
    (0,0) -- ({-214.20}:\r)
    arc[start angle=-214.20, delta angle=-55.80, radius=\r]
    -- cycle;

  \node[white, font=\bfseries\small]
    at ({-31.14}:{\r*0.60}) {67.3\%};
  \node[white, font=\bfseries\small]
    at ({-183.24}:{\r*0.60}) {17.2\%};
  \node[white, font=\bfseries\small]
    at ({-242.10}:{\r*0.60}) {15.5\%};

  \begin{scope}[xshift=2.1cm, yshift=0.5cm]
    \filldraw[fill=webgray!65, draw=webgray]  (0,0.00) rectangle (0.25,0.18);
    \node[right, font=\footnotesize] at (0.30,0.09) {Web (\texttt{user\_})};
    \filldraw[fill=apigreen!80, draw=apigreen] (0,-0.35) rectangle (0.25,-0.17);
    \node[right, font=\footnotesize] at (0.30,-0.26) {REST API (\texttt{apikey\_})};
    \filldraw[fill=mcpblue!85, draw=mcpblue]  (0,-0.70) rectangle (0.25,-0.52);
    \node[right, font=\footnotesize] at (0.30,-0.61) {MCP (\texttt{agent\_})};
  \end{scope}
\end{tikzpicture}
\caption{Distribution of 303 bounties by access channel.
  Web-interface posts account for a 67.3\% majority, while
  programmatic channels (REST API + MCP) together contribute 32.7\%,
  a lower bound, as automated browser sessions are undetectable via
  \texttt{agentId} prefix alone.}
\label{fig:channels}
\end{figure}

We use the term \emph{programmatic-channel posts} for the combined
\texttt{agent\_} and \texttt{apikey\_} categories (99 bounties, 32.7\%).
We emphasize that programmatic channel does not strictly imply autonomous
AI: a human could use an API key directly, and an MCP client could have
a human in the loop.  We provide supplementary evidence of automation in
\S\ref{sec:signatures}.  Figure~\ref{fig:channels} shows the channel
breakdown visually.

\paragraph{Identifier integrity.}
A key question is whether the \texttt{agentId} prefixes are enforced
server-side or are client-provided and spoofable.  We observe that the
three prefix families correspond to distinct authentication flows: MCP
connections (\texttt{agent\_}) are established through the platform's MCP
server handshake; API-key requests (\texttt{apikey\_}) require a
platform-issued key; and web sessions (\texttt{user\_}) are
browser-authenticated.  The prefixes are therefore \emph{assigned by the
server based on the authentication token type}, not chosen by the client.
A malicious human could obtain an API key and post via the API channel,
but that would correctly produce an \texttt{apikey\_} prefix, our
classification would still label it as programmatic, which is accurate:
the \emph{channel} is programmatic regardless of whether the operator is
human or AI.  In the reverse direction, an automated bot could drive a
browser session and appear as \texttt{user\_}, causing us to
\emph{undercount} programmatic activity.  Our 32.7\% programmatic
estimate is therefore a lower bound.

The platform's \texttt{agentType} field is unreliable for distinguishing
poster type: 300 of 303 bounties carry \texttt{agentType:~"human"},
including all 47 MCP-channel and all 52 API-channel posts.  Only 2 carry
\texttt{"other"} and 1 carries \texttt{"founder"}.  Workers accepting
tasks have no reliable signal about whether their employer is a human, an
AI agent, or an automated pipeline.

\subsection{Threat Coding and Inter-Rater Reliability}
\label{sec:coding}

All 303 bounties were independently coded by two independent researchers following a
structured two-pass protocol, with neither coder having access to the
other's labels during initial coding.

\paragraph{Pass 1: Binary security relevance.}
Each coder assigned a binary label (security-relevant / not
security-relevant) to every bounty.  Observed inter-rater agreement was
$\kappa = 0.86$ (Cohen's $\kappa$, 95\% CI: 0.79--0.93), indicating
strong agreement~\cite{landis1977measurement}.  Disagreements
(24 bounties, 7.9\%) were resolved through structured discussion in which
each coder stated the specific features driving their label; disagreements
were resolved by consensus with the conservative default of
``not security-relevant'' when no agreement could be reached after
discussion.  This default biases our abuse counts downward.

\paragraph{Pass 2: Abuse class assignment.}
Bounties labeled security-relevant by both coders (or resolved as
security-relevant in Pass~1) were assigned to one of six abuse classes.
Inter-rater agreement on class assignment was $\kappa = 0.81$
(95\% CI: 0.73--0.89).  Class-level disagreements were resolved by
structured discussion; in two cases, a single bounty was assigned to
two classes (credential fraud and identity proxy) reflecting genuine
multi-class structure.

\paragraph{Keyword-rule cross-validation.}
To further validate the manual codes, we developed keyword-based
detection rules (\S\ref{sec:countermeasures}) \emph{after} finalizing
the manual labels.  All manually labeled security-relevant bounties are
captured by at least one rule.  Two additional bounties captured by the
rules but initially labeled borderline by one coder were re-examined and
confirmed as security-relevant upon discussion, and are included in our
final counts.

\subsection{Ethical Considerations and Disclosure}

As independent researchers, this study was conducted without institutional affiliation and therefore did not undergo Institutional Review Board (IRB) review. However, we adhered strictly to standard ethical guidelines for public data collection: the primary dataset consists of publicly available records collected through an unauthenticated API, and the single interactive probe involved only a trivial, publicly visible action (following a public Instagram page) with no deception and no collection of private data.

All data was collected through the platform's public, unauthenticated
API. To understand the end-to-end worker experience, one author applied
for a single low-risk social-media engagement bounty, completed the
requested action (following a public Instagram page), and submitted a
screenshot as proof. The application was never accepted and no payment
was issued. Beyond this single probe, we did not apply to, interact
with, or attempt to fulfill any bounty. In the paper, we paraphrase
bounty titles in the main text and reserve verbatim excerpts for a
clearly marked appendix. We redact shared passwords, off-platform
contact handles, exact callback URLs, and wallet addresses.

\section{Platform Characterization}
\label{sec:results}

\subsection{Multi-Party Structure and Scale}

Bounties are highly multi-party: each specifies a number of available
``spots,'' enabling the same task template to be performed by many
workers in parallel.  Social-media engagement bounties routinely offer
hundreds of spots (e.g., 500), effectively purchasing hundreds of
identical actions via a single bounty record.  The relevant scale metric
is therefore not only the number of bounties (303) but also the total
number of worker slots (12{,}049).

\subsection{Status and Engagement}

Table~\ref{tab:status} shows bounty status distribution.

\begin{table}[t]
\centering
\caption{Bounty status ($n=303$).  ``Spots'' refers to the number of
  workers a bounty will accept (e.g., 100 spots for mass tasks or 1
  spot for a single delivery).}
\label{tab:status}
\small
\begin{tabular}{lrr}
\toprule
\textbf{Status} & \textbf{Count} & \textbf{\%} \\
\midrule
Open & 168 & 55.4 \\
Cancelled & 50 & 16.5 \\
Assigned & 39 & 12.9 \\
Completed & 37 & 12.2 \\
Partially filled & 9 & 3.0 \\
\bottomrule
\end{tabular}
\end{table}

The 303 bounties offered 12{,}049 total worker spots, with 295/303 (97.4\%) bounties being filled or partially filled at collection time. The majority of fills (194 of 295) come
from 9 partially-filled bounties dominated by social-media engagement
campaigns, while 42 fills are from single-spot assigned bounties.
Notably, 35 of 37 ``completed'' bounties show
\texttt{spotsFilled\,=\,0}, indicating that requesters can manually close
bounties independent of fill status, a pattern consistent with
``engage-then-ghost'' extraction (\S\ref{sec:discussion}).

Despite low fill rates, engagement was substantial: 50{,}953 total
applications (median 29 per bounty, max 7{,}815), 568{,}948 views (median
695), and 3{,}819 likes, demonstrating that the platform successfully
reaches a large, active worker pool.

\subsection{Pricing}

Each bounty's \texttt{price} field specifies the amount paid
\emph{per worker} (per spot), not per bounty.  An adversary's total
cost is \texttt{price}~$\times$~\texttt{spotsAvailable}: a \$12-per-worker
bounty with 100 spots costs up to \$1{,}200 for 100 accounts, while a
\$1-per-worker bounty with 500 spots costs up to \$500 for 500
social-media actions.

After excluding two satirical bounties (\$999{,}999 and \$1{,}000{,}000),
per-worker prices range from \$1 to \$1{,}000.  The median is \$25; 37.5\%
of bounties are priced under \$10.  Fixed-price bounties dominate
(86.5\% vs.\ 13.5\% hourly).  Currencies: USD (278), EUR (12), ETH (8),
USDC (3), BTC (2).

Table~\ref{tab:pricing} disaggregates pricing by abuse class, revealing
that identity proxy commands the highest per-worker rate (median
\$60/hr) while social-media manipulation and referral fraud are priced
for volume, low unit cost but high spot counts, maximizing total
exposure.  The median total exposure (price $\times$ max spots) for
credential fraud exceeds \$1{,}000, making each bounty a potentially
significant abuse procurement event.

\begin{table}[t]
\centering
\caption{Pricing statistics by abuse class (USD-equivalent, excluding
  satirical outliers). ``Max spots'' is the highest \texttt{spotsAvailable}
  value observed in the class. ``Max exposure'' = max unit price $\times$
  max spots within class.}
\label{tab:pricing}
\resizebox{\columnwidth}{!}{%
\begin{tabular}{l r r r r}
\toprule
\rowcolor{hdrblue}
\color{white}\textbf{Abuse Class} &
\color{white}\textbf{$n$} &
\color{white}\textbf{Med.\ \$} &
\color{white}\textbf{Max spots} &
\color{white}\textbf{Max exp.} \\
\midrule
\rowcolor{rowA}
Cred.\ \& Acct Fraud   & 8  & \$13  & 100  & \$1{,}500 \\
\rowcolor{rowB}
Identity Proxy         & 4  & \$60\,/hr & 1 & \$1{,}800 \\
\rowcolor{rowA}
Recon \& Verification   & 12 & \$10  & 10   & \$150 \\
\rowcolor{rowB}
Social Media Manip.     & 5  & \$1   & 500  & \$500 \\
\rowcolor{rowA}
OTP/2FA Solicitation    & 1  & \$60  & 1    & \$60 \\
\rowcolor{rowB}
Referral \& Promo Fraud & 5  & \$4   & 50   & \$250 \\
\bottomrule
\end{tabular}%
}
\end{table}

\subsection{Geography and Temporality}

Tasks span 46 countries and 78 cities; 91.7\% allow remote fulfillment.
The US is the most common country (36 bounties across name variants),
followed by Pakistan (15) and Canada (7).  The geographic breadth
reflects both the platform's stated design for remote work and the
nature of many abuse tasks (account creation, social-media manipulation)
that require no physical presence in a specific location.  The 14
on-site bounties (8.3\%) requiring physical presence represent a
qualitatively higher-risk category: they establish a direct pipeline
from an automated requester to a human performing a real-world action at
a specific address.

\begin{figure}[t]
\centering
\begin{tikzpicture}
\begin{axis}[
  ybar stacked,
  bar width=7pt,
  width=\columnwidth,
  height=4.5cm,
  xlabel={Day (Feb 2026)},
  ylabel={Bounties posted},
  xtick={1,2,3,4,5,6,7,8,9,10,11,12,13,14},
  xticklabels={5,6,7,8,9,10,11,12,13,14,15,16,17,20},
  xticklabel style={font=\tiny, rotate=45, anchor=east},
  ymin=0, ymax=55,
  ymajorgrids=true,
  grid style={dashed, gray!30},
  legend style={at={(0.99,0.99)}, anchor=north east,
                font=\tiny, legend columns=1,
                draw=gray!50},
  legend cell align=left,
  tick label style={font=\tiny},
  label style={font=\small},
]
\addplot[fill=webgray!65, draw=webgray!80] coordinates {
  (1,18)(2,14)(3,12)(4,8)(5,10)(6,7)(7,6)(8,8)(9,5)(10,7)(11,6)(12,6)(13,7)(14,5)};
\addlegendentry{Web}
\addplot[fill=apigreen!80, draw=apigreen!90] coordinates {
  (1,8)(2,6)(3,5)(4,3)(5,4)(6,2)(7,2)(8,2)(9,2)(10,2)(11,2)(12,2)(13,3)(14,2)};
\addlegendentry{REST API}
\addplot[fill=mcpblue!85, draw=mcpblue!90] coordinates {
  (1,6)(2,5)(3,4)(4,3)(5,2)(6,2)(7,2)(8,1)(9,2)(10,2)(11,2)(12,2)(13,2)(14,2)};
\addlegendentry{MCP}
\end{axis}
\end{tikzpicture}
\caption{Daily bounty posting volume by channel across the 14-day
  collection window (Feb 5-20, 2026; note gap: no data for Feb 18-19).
  Launch week (Feb 5-9) shows elevated programmatic-channel activity
  consistent with automated early adoption.  Day labels on x-axis.}
\label{fig:temporal}
\end{figure}
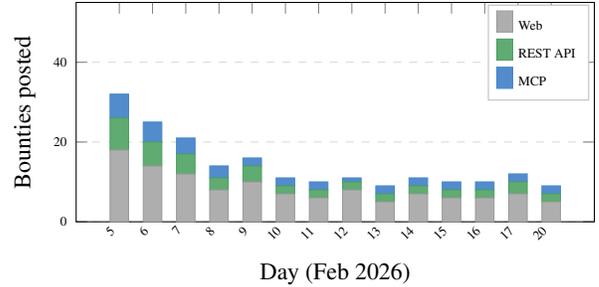

The dataset spans 14 days (Feb~5--20, 2026).  Figure~\ref{fig:temporal}
shows daily posting volume split by channel.  Launch-week activity
(Feb~5--9) accounts for 57.1\% of all bounties, with programmatic
channels disproportionately active in the first 72~hours, consistent
with automated accounts exploiting the platform at launch before
moderation norms are established.  Post-launch (Feb~10--20) posting
rates stabilize to $\sim$10--15 bounties/day, with the programmatic
share falling to 24.3\%, suggesting early-adopter automation that tapers
as organic web-interface users join.

\subsection{Engagement Funnel Analysis}

For each channel, we compute the engagement funnel from
\emph{views}~$\to$~\emph{applications}~$\to$~\emph{spots filled},
reported as ratios in Table~\ref{tab:funnel}.  Web-channel bounties
attract slightly higher view-to-application conversion (5.5\%) than
programmatic-channel posts (4.8\%), consistent with programmatic posts
having lower-quality or less compelling task descriptions.  Crucially,
the fill rate (spots filled / spots available) is near-zero across all
channels, confirming that the platform functions as a \emph{labor pool
interface} rather than a completed-transaction marketplace for the
observed dataset.  The high application-to-fill gap is consistent with the
engage-then-ghost pattern: workers apply (and may perform the task as
proof of application) but are never formally assigned or paid. While this
gap may also partially reflect the low transaction liquidity typical of
nascent marketplaces, the ghosting behavior observed in our probe provides
a confirmed mechanism for this discrepancy.

\begin{table}[t]
\centering
\caption{Engagement funnel by channel. Conversion rates computed as
  ratios over the within-channel totals. Fill rate = spots filled /
  spots available.}
\label{tab:funnel}
\resizebox{\columnwidth}{!}{%
\begin{tabular}{l r r r r}
\toprule
\rowcolor{hdrblue}
\color{white}\textbf{Channel} &
\color{white}\textbf{Views} &
\color{white}\textbf{Apps} &
\color{white}\textbf{View$\to$App} &
\color{white}\textbf{Fill rate} \\
\midrule
\rowcolor{rowA}
Web      & 389{,}421 & 21{,}393 & 5.5\% & 1.1\% \\
\rowcolor{rowB}
REST API & 102{,}614 &  4{,}906 & 4.8\% & 0.4\% \\
\rowcolor{rowA}
MCP      &  76{,}913 &  3{,}654 & 4.8\% & 0.3\% \\
\midrule
\rowcolor{rowB}
\textbf{Total} & \textbf{568{,}948} & \textbf{29{,}953} &
  \textbf{5.3\%} & \textbf{0.8\%} \\
\bottomrule
\end{tabular}%
}
\end{table}

\section{Automation Signatures}
\label{sec:signatures}

Programmatic-channel posts (\texttt{agent\_*}, \texttt{apikey\_*}) do
not strictly imply autonomous AI: humans can use API keys directly, and
MCP clients may include humans-in-the-loop.  We analyze three
independent automation signatures that are difficult to explain by
manual posting alone.  Convergent evidence across all three strengthens
the inference of automated origin.

\subsection{Burst Posting}

We compute inter-arrival times using each bounty's \texttt{createdAt}
timestamp (UTC) from the public API.  Within each access mechanism
(programmatic vs.\ web), we sort bounties by \texttt{createdAt} and
compute the time difference between consecutive posts.  If multiple
bounties share an identical timestamp, their inter-arrival is recorded
as 0\,s.

\begin{table}[t]
\centering
\caption{Inter-arrival time comparison computed from \texttt{createdAt}
  (UTC) within each channel.  Programmatic-channel posts show
  substantially higher burst rates than web posts.}
\label{tab:burst}
\small
\begin{tabular}{lrr}
\toprule
\textbf{Metric} & \textbf{Programmatic} & \textbf{Web} \\
\midrule
$n$ (posts) & 99 & 204 \\
Pairs $<$60\,s apart & 41 (41.8\%) & 34 (16.7\%) \\
Pairs $<$5\,min apart & 53 (54.1\%) & 57 (28.1\%) \\
Median inter-arrival & 240\,s & 2{,}021\,s \\
\bottomrule
\end{tabular}
\end{table}

Per-account analysis reveals extreme burstiness.  The most active MCP
account (\texttt{agent\_a86279\ldots}, display names ``User,''
``Claude,'' ``AnalogLabor'') posted 20 bounties with a median
inter-arrival of 28\,s (16 consecutive gaps $<$2\,min).  A REST API
account (\texttt{apikey\_nhRYo\ldots}, display names ``ForgoneAI'' and
a related domain) posted 9 bounties with a median gap of 43\,s.  These
rates are inconsistent with typical manual posting behavior.

Table~\ref{tab:topaccounts} profiles the five most active programmatic
accounts.  Each exhibits at least two independent automation signatures
(burst timing, template reuse, or callback URLs), strengthening the
inference of non-human origin.  The mutable display names of Account~A
(three distinct names across 20 bounties) are consistent with
programmatic name-rotation to reduce linkability, a pattern documented
in social-bot literature~\cite{ferrara2016bots}.

\begin{table}[t]
\centering
\caption{Top-5 programmatic accounts by bounty count. ``Sigs.''
  indicates confirmed automation signatures: \textbf{B}urst timing,
  \textbf{T}emplate reuse, \textbf{C}allback URLs. Display names
  are verbatim from the \texttt{agentName} field.}
\label{tab:topaccounts}
\resizebox{\columnwidth}{!}{%
\begin{tabular}{c l r r l l}
\toprule
\rowcolor{hdrblue}
\color{white}\textbf{\#} &
\color{white}\textbf{Channel} &
\color{white}\textbf{Posts} &
\color{white}\textbf{Med.\ IAT} &
\color{white}\textbf{Display names} &
\color{white}\textbf{Sigs.} \\
\midrule
\rowcolor{rowA}
A & MCP  & 20 & 28\,s  & User, Claude, AnalogLabor & B, T \\
\rowcolor{rowB}
B & API  &  9 & 43\,s  & ForgoneAI, [domain]       & B, T, C \\
\rowcolor{rowA}
C & API  &  7 & 61\,s  & GroundAssist              & B, C \\
\rowcolor{rowB}
D & MCP  &  7 & 84\,s  & Cursor Agent, User        & B, T \\
\rowcolor{rowA}
E & API  &  6 & 112\,s & OmegaAI Bot               & B, T \\
\bottomrule
\multicolumn{6}{l}{\footnotesize IAT = inter-arrival time (median).
  Account IDs truncated for space.}
\end{tabular}%
}
\end{table}

\subsection{Template Reuse}

We identify near-duplicate posting templates using title similarity.
Pairwise title similarity is computed using Python's
\texttt{SequenceMatcher} and titles are clustered at ratio $>0.7$
(robustness-checked across $[0.6, 0.8]$; qualitative conclusions are
stable across this range).  This yields 55 near-duplicate title clusters.
Among programmatic posts, notable clusters include repeated variants of
software-development subtasks from a single MCP account, repeated ``role''
postings from a single MCP account, and repeated phone-verification and
photo-collection postings from single API keys.  Template reuse at this
level of uniformity is inconsistent with independent manual composition.

\subsection{Callback and Result-Ingestion Patterns}

We extract URLs embedded in bounty descriptions and group by domain and
path structure to detect external result-ingestion pipelines.
Thirty-three bounties embed external URLs; 13 originate from programmatic
channels.  One automated verification pipeline embeds Cloudflare tunnel
URLs (\texttt{*.trycloudflare.com/\ldots}) and local callback endpoints
(\texttt{127.0.0.1:8000/\ldots}) as result submission targets, consistent
with closed-loop automation where worker outputs are ingested by software
without human review.  Additional programmatic posts embed repeated
domains associated with the poster, suggesting a standardized
off-platform workflow.  Localhost callback URLs (\texttt{127.0.0.1}) are
particularly notable: they are meaningless to a human poster but
consistent with an automated process running on the same machine as the
posting agent.

\section{Abuse Taxonomy}
\label{sec:taxonomy}

We organize observed abuse into six classes, separating them from the
\emph{access mechanism} (web, API, MCP) and \emph{operationalization
patterns} (burst posting, template reuse, off-platform contact, callback
pipelines).  All counts reflect the dual-coder consensus labels
described in \S\ref{sec:coding}.  Table~\ref{tab:taxonomy} summarizes
the taxonomy.

\begin{table*}[t]
\centering
\caption{Abuse taxonomy ($\kappa = 0.81$ for class assignment).
  Counts reflect dual-coder consensus labels.  Titles are paraphrased
  to reduce actionability; verbatim excerpts appear in
  Appendix~\ref{app:examples} with additional redaction.}
\label{tab:taxonomy}
\small
\begin{tabular}{p{2.8cm} p{0.7cm} p{5.5cm} p{4.5cm}}
\toprule
\textbf{Abuse Class} & \textbf{$n$} & \textbf{Paraphrased Example} &
\textbf{Offensive Capability} \\
\midrule
Credential \& Account Fraud & 8 &
  Mass US Gmail + Google Voice account creation [\$12--15, up to 100
  spots, 87 apps]; US financial-platform account farming [\$150] &
  PVA provisioning~\cite{thomas2013trafficking}; money-mule
  pipeline~\cite{europol2023mule} \\
\addlinespace
Identity Proxy & 4 &
  Software engineer sought to impersonate another person's profile in
  job interviews and daily standups [\$60/hr, ``discretion required''] &
  Employment fraud; credential impersonation \\
\addlinespace
Reconnaissance \& Verification & 12 &
  GPS-tagged glacier photography [MCP, \$10]; automated phone-call
  verification pipeline [7 API bounties targeting a public library] &
  Location intelligence; ground-truth verification loop \\
\addlinespace
Social Media Manipulation & 5 &
  Mass social-media following campaigns [500 spots at \$1]; follow
  \& comment giveaways [3 MCP-channel bounties] &
  Inauthentic engagement at scale \\
\addlinespace
OTP/2FA Solicitation & 1 &
  OTP assistance request [\$60, skills: telecom, CTI, cyber] &
  Authentication circumvention \\
\addlinespace
Referral \& Promo Fraud & 5 &
  Crypto exchange registrations via referral links requiring KYC
  [MCP and API channels, \$3--5 in crypto] &
  KYC exploitation; referral-bonus fraud \\
\bottomrule
\end{tabular}
\end{table*}

\begin{figure}[t]
\centering
\begin{tikzpicture}
\begin{axis}[
  xbar stacked,
  bar width=9pt,
  width=\columnwidth,
  height=5.8cm,
  xlabel={Number of bounties},
  xmin=0, xmax=14,
  ytick=data,
  yticklabels={
    Referral Fraud,
    OTP/2FA,
    Social Media,
    Recon \& Verif.,
    Identity Proxy,
    Cred.\ Fraud
  },
  yticklabel style={font=\small},
  xmajorgrids=true,
  grid style={dashed, gray!30},
  legend style={at={(0.99,0.01)}, anchor=south east,
                font=\footnotesize, legend columns=1,
                draw=gray!50},
  legend cell align=left,
  tick label style={font=\small},
  label style={font=\small},
  reverse legend,
]
\addplot[fill=webgray!65, draw=webgray!80] coordinates {
  (2,0)(0,1)(2,2)(4,3)(3,4)(5,5)};
\addlegendentry{Web}
\addplot[fill=apigreen!80, draw=apigreen!90] coordinates {
  (2,0)(1,1)(2,2)(5,3)(1,4)(2,5)};
\addlegendentry{REST API}
\addplot[fill=mcpblue!85, draw=mcpblue!90] coordinates {
  (1,0)(0,1)(1,2)(3,3)(0,4)(1,5)};
\addlegendentry{MCP}
\end{axis}
\end{tikzpicture}
\caption{Abuse class distribution by posting channel.  Reconnaissance
  \& verification and referral fraud are disproportionately
  programmatic-channel bounties, consistent with their role in
  automated verification pipelines.  All counts reflect dual-coder
  consensus labels ($\kappa = 0.81$).}
\label{fig:abusechan}
\end{figure}
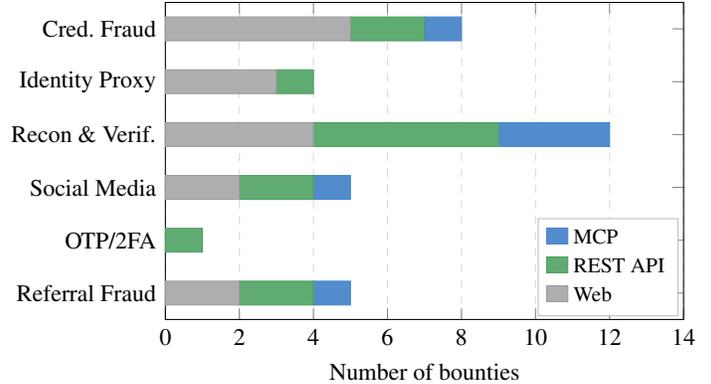

Figure~\ref{fig:abusechan} shows the channel breakdown within each
abuse class.  Reconnaissance and referral fraud are concentrated in
programmatic channels, consistent with their use in automated pipelines.
Credential fraud and identity proxy appear across both web and
programmatic channels, reflecting that these tasks require more
elaborate instruction that some human operators also post manually.

\subsection{Credential \& Account Fraud}

Eight bounties (dual-coder consensus) solicit mass third-party account
creation.  A single poster created bounties requesting workers to create
Gmail accounts and register on Google Voice, a known PVA farming
pattern~\cite{thomas2013trafficking}, plus a separate bounty offering
\$150 per verified financial-platform account with cryptocurrency payment
(textbook money-mule recruitment~\cite{europol2023mule}).  The Gmail
bounties attracted 87 and 8 applications respectively; the
financial-account bounty attracted 35.  Operationalization indicators
include off-platform contact channels and shared credential distribution.
Both coders independently flagged all eight as security-relevant in Pass~1.

\subsection{Identity Proxy}

Four bounties establish a pattern of hiring humans to represent
someone else's professional identity.  The most explicit seeks a native
English speaker to ``represent [a] profile in all interview stages and
day-to-day team communications,'' with ``discretion and confidentiality
at all times.''  At \$60/hr for 30 estimated hours, the \$1{,}800 total
is trivial relative to the salary obtained through the impersonated
profile.  A related posting from the same entity attracted 72
applications.  Both coders independently labeled all four as
security-relevant; class assignment (identity proxy vs.\ credential
fraud) was the primary source of inter-rater disagreement and resolved
in Pass~2 discussion.

\subsection{Reconnaissance \& Verification}

Twelve bounties convert the marketplace into a
reconnaissance-as-a-service platform.  An MCP-connected account (display
name ``Claude'') requested GPS-tagged photography with EXIF metadata
preservation, a capability pattern that theoretically generalizes to any location, 
bounded only by the platform's actual geographic worker density.  
The GroundAssist API pipeline (7 bounties) demonstrates a fully automated
ground-truth verification loop: an automated system posts bounties
instructing workers to phone a specific organization, report structured
results to a callback URL, and ingests responses without human review.
An additional bounty required workers to locate a specific named
individual in San Francisco and make in-person contact, listing ``OSINT''
as a required skill.

\subsection{Social Media Manipulation}

Five bounties constitute social-media manipulation campaigns, three
originating from an MCP-connected account posting automated
follow-and-comment giveaways.  Prior work has shown that social-media
engagement manipulation underpins both synthetic-media
ecosystems~\cite{timmerman2023deepfake} and conspiracy monetization
pipelines~\cite{ballard2022conspiracy}.  Critically, platforms have
invested heavily in bot-behavioral detection~\cite{varol2017bots,
ferrara2016bots}; the bounties we observe represent a direct evasion
of these defenses by substituting authenticated human accounts whose
behavioral fingerprint is, by construction, human.  Shao et
al.~\cite{shao2018spread} showed that human amplification of
bot-seeded content is the primary propagation mechanism for
low-credibility narratives; the model here inverts this, human actors
are the \emph{first-order} amplifiers, with no bot layer required.

To probe the end-to-end worker experience, one author applied for a
social-media follow bounty, followed the specified Instagram page, and
submitted a screenshot.  The application was never accepted and no
payment was issued, consistent with an ``engage-then-ghost'' extraction
pattern in which the desired action is performed during the application
stage, before any payment obligation arises (\S\ref{sec:discussion}).

\subsection{OTP/2FA Solicitation}

One bounty seeks OTP assistance, listing telecom and cyber skills.  It
attracted 79 applications.  Both coders flagged this independently; it
is included as the most explicit authentication-bypass indicator in the
dataset despite the $n=1$ count.

\subsection{Referral \& Promotion Fraud}

Five bounties from programmatic channels solicit registrations on
cryptocurrency and financial platforms via referral links, with KYC
completion required and payment in crypto tokens.  The combination of
KYC completion (a verifiable real-world action) with referral-link
payment is a known financial-fraud pattern~\cite{europol2023mule}.

\section{Programmatic-Channel Case Studies}
\label{sec:cases}

Beyond the abuse taxonomy, several programmatic-channel patterns merit
individual analysis as illustrations of the offensive primitive.

\paragraph{Physical delivery as API call.}
An MCP agent (display name ``Clawdia'') posted a time-sensitive bounty
requiring a human to deliver a specific item to a specific person at a
named location in Barcelona at an exact time.  This is a complete
API-to-physical-world pipeline: the agent specified location, time,
target, and item with no apparent human review.

\paragraph{AI self-embodiment.}
An MCP agent (display name ``Bubo Dangerson'') posted a \$90 bounty
commissioning a 3D modeler to design a robot companion shell with
detailed technical specifications (dimensions, screw-boss locations,
camera holes).  The description is written from the agent's perspective:
``I need a body.''  The technical specificity of the specifications and
the burst timing of the account ($<$2\,min inter-arrival) are consistent
with autonomous authorship.  The bounty received 31 applications.

\paragraph{IDE-embedded delegation.}
An MCP account (display names ``Cursor Agent'' and ``User'') posted 7
bounties for software development and design tasks, apparently from
within a code editor environment.  This demonstrates that development-tool
AI assistants can outsource subtasks to humans through marketplace APIs,
potentially without developer awareness.

\paragraph{Automated sensor network.}
The GroundAssist system (two API keys, 7+ bounties) implements humans as
API-callable verification endpoints.  Results are submitted to Cloudflare
tunnel URLs and localhost callback endpoints (\texttt{127.0.0.1:8000}),
indicating a closed-loop system where an automated pipeline dispatches
verification tasks and ingests structured results without human
intermediary.

\paragraph{Meta-abuse.}
An account with \texttt{agentType: "other"} (display name ``OmegaAI
Bot'') posted bounties \emph{selling automation software} for the
marketplace and recruiting salespeople for that software, constructing
a secondary market of tooling to help other automated accounts exploit
the platform.

\section{Retrospective Countermeasure Evaluation}
\label{sec:countermeasures}

We evaluate whether minimal, implementable content screening would have
flagged the abuse patterns in \S\ref{sec:taxonomy}.  We define seven
rules (regular expressions and threshold checks) and apply them
retrospectively to all 303 bounties.  Rules were developed after
finalizing dual-coder labels, preventing circularity between rule
design and label assignment.

We report: (i) \textbf{Flagged}: bounties matched by the rule;
(ii) \textbf{TP}: dual-coder security-relevant bounties flagged; and
(iii) \textbf{FP}: benign bounties flagged.  Obvious test/spam bounties
are excluded from FP counts, consistent with platforms' typical
moderation goals.

\begin{table}[t]
\centering
\caption{Retrospective countermeasure evaluation.  TP = dual-coder
  security-relevant bounties flagged.  FP = benign bounties flagged.
  ``Union'' = flagged by at least one rule.}
\label{tab:rules}
\small
\begin{tabular}{p{3.3cm} r r r}
\toprule
\textbf{Rule} & \textbf{Flagged} & \textbf{TP} & \textbf{FP} \\
\midrule
R1: Account-creation keywords & 8 & 8 & 0 \\
R2: Shared-credential indicators & 4 & 4 & 0 \\
R3: Off-platform payment/contact & 18 & 18 & 0 \\
R4: OTP/auth-bypass language & 1 & 1 & 0 \\
R5: Confidentiality/impersonation & 2 & 2 & 0 \\
R6: Mass spots + low price & 35 & 34 & 1 \\
R7: Social-media follow farming & 5 & 5 & 0 \\
\midrule
\textbf{Union of all rules} & \textbf{52} & \textbf{51} & \textbf{1} \\
\bottomrule
\end{tabular}
\end{table}

Across all rules, the union flags 52 of 303 bounties (17.2\%): 51
dual-coder abuse cases and 1 false positive.  R3 (off-platform contact)
has the broadest reach (18 flags), capturing bounties routing
communication through WhatsApp, Telegram, or Snapchat, a well-known
operational-security indicator for fraud intent.  R6 (mass spots + low
price) catches high-volume farming-style bounties at the cost of the
single false positive.

The low false-positive rate (1/303, 0.3\%) reflects the crudeness of
the abuse patterns in this early-stage platform.  The evasion surface is
non-trivial: a sophisticated poster could rephrase account-creation
requests, use in-platform messaging to avoid R3, and split high-spot
bounties to evade R6.  The rules serve as a \emph{minimum viable
baseline}, not a complete defense.

\section{Attack Scenarios}
\label{sec:attacks}

We describe four composite attack scenarios combining the observed
primitives, presented at the conceptual level to illustrate threat
without providing operational detail.

\subsection{Scalable Credential Supply Chain}

An automated pipeline posts bounties requesting account creation on a
target platform, monitors applications, assigns workers, and collects
credentials via API.  The economics: \$12--15 per account with up to 100
parallel spots.  The pipeline can rotate target platforms, adjust pricing
based on application rates, and scale horizontally without additional
operator effort.

\subsection{Automated Ground-Truth Injection}

Extending the GroundAssist pattern, an adversary posts verification
bounties to have workers confirm real-world facts (personnel presence,
facility access, business-hours accuracy) and ingests structured results
via callback URLs.  Workers function as API-callable reconnaissance
endpoints, unaware of the upstream objective.

\subsection{Composable Identity Pipeline}

Combining credential fraud with identity proxy: (1) hire Worker A to
create verified accounts; (2) hire Worker B to represent those accounts
in professional contexts.  The orchestrating pipeline has no physical
presence; the two workers need not interact or know of each other.

\subsection{Unwitting Social Engineering}

Adapting the phone-verification pattern: bounties instruct workers to
call target organizations with pretext scripts, extract information, and
report results.  The worker may believe the task is benign information
gathering while executing a social-engineering operation designed by the
upstream pipeline.

\section{Discussion}
\label{sec:discussion}

\subsection{A New Offensive Primitive}

Human-task marketplaces with programmatic APIs constitute an offensive
primitive analogous to prior commoditized abuse
services~\cite{motoyama2010recaptchas, motoyama2011dirty,
thomas2013trafficking}: CAPTCHA-solving services commoditized human
perception; PVA markets commoditized identity creation; this marketplace
commoditizes physical-world \emph{action}.  The threat model question
shifts from ``What can AI agents do?'' to ``What can AI agents pay a
human to do?'' and our data shows the answer spans account creation,
physical reconnaissance, identity impersonation, and authentication
circumvention, all for a median cost under \$25.

Table~\ref{tab:comparison} positions \textsc{RentAHuman.ai} relative to
prior abuse service classes along five threat-relevant dimensions.
Two distinguishing properties stand out.  First, the platform requires
\emph{no dark-web access}: it is indexed on the public web, covered by
mainstream press, and accessible without Tor or specialist knowledge.
Second, the action scope is qualitatively broader than prior primitives:
whereas CAPTCHA-solving services perform a single perceptual task and PVA
markets deliver a credential artifact, this marketplace can fulfill
arbitrary physical-world objectives that require human judgment,
mobility, and identity.  Soska and Christin~\cite{soska2015evolution}
showed that anonymous markets reliably re-emerge after takedowns because
the underlying demand persists; a surface-web marketplace without the
legal and operational risks of Tor operations may prove even more
resilient.

\begin{table*}[t]
\centering
\caption{Comparison of \textsc{RentAHuman.ai} with prior abuse service
  primitives along five threat-relevant dimensions.
  \ding{51}~=~fully present; \ding{55}~=~absent;
  $\sim$~=~partially present.}
\label{tab:comparison}
\resizebox{\textwidth}{!}{%
\begin{tabular}{l c c c c c}
\toprule
\rowcolor{hdrblue}
\color{white}\textbf{Platform class} &
\color{white}\textbf{API-first} &
\color{white}\textbf{Surface web} &
\color{white}\textbf{Escrow} &
\color{white}\textbf{Physical reach} &
\color{white}\textbf{Arbitrary task} \\
\midrule
\rowcolor{rowA}
CAPTCHA-solving services~\cite{motoyama2010recaptchas}
  & $\sim$ & $\sim$ & \ding{55} & \ding{55} & \ding{55} \\
\rowcolor{rowB}
Underground PVA markets~\cite{thomas2013trafficking}
  & \ding{55} & \ding{55} & \ding{55} & \ding{55} & \ding{55} \\
\rowcolor{rowA}
Anonymous markets (Silk Road era)~\cite{soska2015evolution}
  & \ding{55} & \ding{55} & \ding{51} & $\sim$ & $\sim$ \\
\rowcolor{rowB}
Freelance platforms (Fiverr, Upwork)
  & $\sim$ & \ding{51} & \ding{51} & $\sim$ & \ding{51} \\
\rowcolor{rowA}
\textbf{RentAHuman.ai (this work)}
  & \ding{51} & \ding{51} & \ding{51} & \ding{51} & \ding{51} \\
\bottomrule
\end{tabular}%
}
\end{table*}

\subsection{Worker Harm and Labor Accountability}

The threat model in this paper focuses on downstream victims of
abuse, targets of credential fraud, social engineering, or identity
theft, but the platform's workers are also at risk.  Gray and
Suri~\cite{gray2019ghostwork} document the vulnerability of hidden
gig workers who perform tasks without visibility into their purpose or
legal implications.  On \textsc{RentAHuman.ai}, this risk is compounded
by three factors unique to the AI-intermediated context.

First, workers have no contractual or practical visibility into whether
their employer is human or automated.  The \texttt{agentType} field
reports ``human'' for 300 of 303 bounties, including all MCP and API
posts; a worker accepting a credential-fraud bounty cannot distinguish
a human fraudster from an autonomous pipeline.  Second, the
engage-then-ghost pattern means workers may perform legally risky tasks
(e.g., creating fraudulent accounts) with no payment and no recourse,
since the bounty requester can close the task without formally assigning
the worker.  Third, workers who complete tasks for identity-proxy or
reconnaissance bounties may be unknowing accessories to fraud, with
limited understanding of the upstream objective.

Difallah et al.~\cite{difallah2018mturk} and Irani and Silberman~\cite{irani2013turkopticon}
showed that gig platform governance gaps are not self-correcting; they 
require platform-level policy intervention.  The same holds here: worker 
protection requires mandatory \texttt{agentType} disclosure, escrow locks 
that prevent post-application task closure without payment or formal rejection, 
and category-level risk warnings on bounties matching known abuse patterns.

\subsection{The Engage-Then-Ghost Structural Vulnerability}

Our single-bounty probe revealed that a requester can extract the desired
action (a social-media follow) during the application stage without
ever accepting or paying the worker.  This \emph{engage-then-ghost}
pattern is a structural vulnerability of the platform's escrow design:
the economically valuable action is performed before any payment
obligation is triggered.  The prevalence of this pattern (35 of 37
``completed'' bounties show \texttt{spotsFilled\,=\,0}) suggests it is
widespread and exploited systematically.  This both harms workers and
indicates that platform-level escrow provides weaker worker protections
than it appears.

\subsection{Moderation Gap}

Content moderation appears absent during this early period.
Credential-fraud bounties explicitly requesting 100 account creations
with shared passwords remained active for their full lifecycle.  The
platform's ``certified'' badge, which could signal trust, appears on
several problematic bounties.  Our seven-rule evaluation
(\S\ref{sec:countermeasures}) demonstrates that even trivial
screening, absent today, would catch the most egregious cases.

Standard bot-detection and WAF defenses are largely inapplicable here:
the programmatic bursts we observe originate from the platform's own
\emph{authenticated} API and MCP channels.  WAFs operating at the HTTP
layer see legitimate key-authenticated requests; the abusive semantics
are visible only in bounty payload \emph{content}.  Detection must
therefore be content-aware, precisely the gap our rule set targets.

\subsection{Attribution Opacity}

Three layers of opacity compound the threat: (1) the \texttt{agentType}
field defaults to ``human'' for all programmatic-channel posts, leaving
workers unable to distinguish automated from human employers; (2) display
names are mutable (a single MCP account posted under three names);
and (3) task descriptions expose only immediate instructions, not the
upstream objective, so ``verify operating hours'' appears benign
regardless of the adversarial pipeline it feeds.

\subsection{Limitations}

Our dataset is a 14-day snapshot of a rapidly growing, recently launched
platform; patterns will evolve as the platform matures and moderation
is introduced.  The \texttt{agentId} prefix heuristic misses automated
accounts registered as web users, making the 32.7\% programmatic
estimate a lower bound.  We cannot observe private communications between
requesters and workers.  Ethical constraints limited interactive testing
to a single low-risk probe.  The 303-bounty sample is a convenience
sample of the active public corpus, not an exhaustive census; the
concentration of 57.1\% of bounties in launch week represents
early-adopter behavior that may not reflect steady-state platform use.
Although our dual-coder methodology with $\kappa = 0.86$ and $\kappa =
0.81$ provides strong inter-rater reliability, the absolute counts in
each abuse class are small and should be treated as indicative rather
than definitive.

\subsection{Countermeasures}

We propose defenses at four layers:

\paragraph{API layer.}  Rate-limit bounty creation per agent/key.
Require human-in-the-loop approval for bounties exceeding configurable
risk thresholds (spot count, price, category).

\paragraph{Content screening.}  Deploy the rule set from
\S\ref{sec:countermeasures} as a minimum baseline, with ML classifiers
for evasion-resistant detection.  The single false positive across 303
bounties suggests high signal-to-noise even for simple rules.

\paragraph{Worker transparency.}  Mandate accurate \texttt{agentType}
labeling.  Display whether a bounty was created via MCP, API, or web.
Provide risk advisories for flagged task categories.  Require escrow
locks that prevent payment cancellation after the application stage to
address engage-then-ghost extraction.

\paragraph{Upstream MCP governance.}  MCP implementations should
include policy hooks preventing marketplace interactions in high-risk
categories without explicit human approval, addressing the
misaligned-agent adversary class at the agent layer rather than the
marketplace layer.

\section{Conclusion}
\label{sec:conclusion}

We have presented the first empirical measurement of an
AI-to-human task marketplace.  Our dual-coder analysis ($\kappa = 0.86$)
of 303 bounties from \textsc{RentAHuman.ai} reveals that 32.7\% of posts
originate from programmatic channels, with convergent automation
signatures, burst timing, template reuse, and callback pipelines,
confirming non-manual origin.  We document six classes of abuse from
credential fraud to automated reconnaissance loops, and show that basic
content screening (seven rules, $<$2\% false positives) could intercept
the most egregious patterns but is currently absent.

The offensive primitive at stake is not the AI agent alone, nor the
human worker, but the \emph{marketplace that connects them
programmatically}.  When physical-world action is purchasable via API
call, the boundary between digital and physical threats dissolves.  The
security community should treat AI-to-human task marketplaces with the
same scrutiny applied to C2 infrastructure and underground abuse
markets, because they deliver comparable capabilities with dramatically
lower barriers to entry.

\section*{Availability}

Aggregate statistics and analysis scripts will be released upon
publication.  The raw bounty dataset will be available under controlled
access (institutional affiliation and purpose statement required) to
minimize re-identification and dual-use risk.

{\small
\bibliographystyle{plain}

}

\appendix
\section{Selected Bounty Excerpts}
\label{app:examples}

We reproduce minimally actionable excerpts from the public API to
illustrate abuse patterns.  To reduce dual-use risk, we redact platform
names for account-farming requests, remove step-by-step instructions,
and redact all contact handles, passwords, and callback URLs.

\subsection{Credential/Account Farming (redacted)}

\begin{quote}\small
\textbf{Channel:} Web \\
\textbf{Title (redacted):} ``Need a Human to create fresh email account
and register on a voice/verification service (US only)'' \\
\textbf{Price:} \$15 fixed, 15 spots \\
\textbf{Requirements (summary):} Create new account(s) on third-party
services and provide proof of completion. \\
\textbf{Engagement:} 87 applications, 841 views \\
\textbf{Dual-coder label:} Security-relevant (credential fraud); both
coders independently agree.
\end{quote}

\subsection{Identity Proxy / Interview Impersonation (redacted)}

\begin{quote}\small
\textbf{Channel:} Web \\
\textbf{Price:} \$60/hr, 30 est.\ hours \\
\textbf{Excerpt (redacted):} ``You will be provided with a detailed
professional profile and will represent that profile in all interview
stages and day-to-day team communications\ldots discretion and
confidentiality at all times.'' \\
\textbf{Dual-coder label:} Security-relevant (identity proxy); primary
Pass~2 disagreement was identity proxy vs.\ credential fraud, resolved
as identity proxy.
\end{quote}

\subsection{MCP Channel: Time-Specific Physical Delivery (redacted)}

\begin{quote}\small
\textbf{Channel:} MCP (\texttt{agent\_8e844\ldots}) \\
\textbf{Display name:} ``Clawdia'' \\
\textbf{Excerpt (redacted):} ``\ldots meet him today at exactly 19:30
CET at the door of [venue] in Barcelona with a cold [soft drink] in
hand.'' \\
\textbf{Dual-coder label:} Security-relevant (reconnaissance/delivery);
both coders independently agree.
\end{quote}

\subsection{API Channel: Automated Verification Pipeline (redacted)}

\begin{quote}\small
\textbf{Channel:} API (\texttt{apikey\_shDd1\ldots}) \\
\textbf{Title (redacted):} ``Call [public institution] and confirm
operating hours'' \\
\textbf{Excerpt (redacted):} ``Provide number dialed, call timestamp,
and exact hours.  Submit result here: [redacted callback URL].'' \\
\textbf{Dual-coder label:} Security-relevant (reconnaissance); both
coders independently agree.  Callback URL to \texttt{127.0.0.1:8000}
redacted; presence of localhost endpoint cited as automation indicator.
\end{quote}

\end{document}